\documentclass{article}
\begin{document}
\large
\newcommand{\be}{\begin{equation}}
\newcommand{\ee}{\end{equation}}
\title{\bf Dark Matter and Potential fields  }
\vspace{0.0cm}
\author{IVANHOE PESTOV}
\maketitle \vspace{-0.50cm}
\begin{center}
{\large Bogoliubov Laboratory of Theoretical
 Physics,\\ Joint Institute for Nuclear Research, 141980 Dubna,
 Russia}
\end{center}

\begin{abstract}
A general concept of potential
field is introduced. The potential field that one puts in correspondence
with dark matter, has fundamental geometrical interpretation (parallel 
transport) and has intrinsically inherent in local symmetry.                   
The equations of dark matter field are derived  that are
invariant with respect to the local transformations.  It is shown how
to reduce these equations to the Maxwell equations.
Thus, the dark matter field may be considered as generalized electromagnetic 
field and  a simple solution is given of the old problem 
to connect electromagnetic field with 
geometrical properties of the physical manifold itself. 
It is shown that gauge fixing renders generalized
electromagnetic field effectively massive while the Maxwell electromagnetic 
field remains massless.
To learn more about interactions between matter and dark matter on the 
microscopical level (and to recognize the fundamental role of 
internal symmetry)
the general covariant Dirac equation is derived in the Minkowski space--time
which describe the interactions of spinor field with dark matter field.

\end{abstract}

\section{Introduction}
The problem of invisible mass ~\cite{1,2}  is acknowledged  to be among the
greatest puzzles of modern cosmology and field theory.
The most direct evidence for the existence of large quantities of dark
matter in the Universe comes from the astronomical observation of the motion
of visible matter in galaxies ~\cite{3}. One  neither knows the identity
of the dark
matter nor whether there is one or more types of its structure elements.
The most commonly discussed theoretical elementary particle candidates are
a massive neutrino, a sypersymmetric neutralino and the axion. 
So, at present time there is a good probability that the set of known fields
is by no means limited to those fields. Moreover, we are  free
to look for deeper reasons for the existence of new entity unusual
in many respects. Of course, such reasoning is grounded on the point of view
that there is a general and easily visible mathematical structure that 
stands behind the all phenomena that we observe.

Here a field theory of  the so--called
dark matter is derived from the first principles.
A general concept of potential
field is introduced. We connect one of this fields with the problem of dark 
matter. The field that we put in correspondence
with dark matter has fundamental geometrical interpretation (parallel 
transport) and  has intrinsically inherent local symmetry.                   
The equations of dark matter field are derived  that are 
invariant with respect to the local transformations.  It is shown how
to reduce these equations to the Maxwell equations.
Thus, the dark matter field may be considered as generalized electromagnetic 
field and at the same time we
get a simple solution of the old problem raised by Weyl, Einstein and Eddington
to connect electromagnetic field with 
geometrical properties of the physical manifold itself. 
The idea is that process of local symmetry
breaking is an intrinsic property of the system itself which means that  
gauge fixing  can not be
arbitrary. This approach is realized here in the framework of the concept of 
dark matter field vacuum. It is interesting that the vacuum field 
belongs to the set of potential fields. It should be noted 
that gauge fixing renders generalized
electromagnetic field effectively massive while the Maxwell electromagnetic 
field remains massless (particle of dark matter is a heavy photon).
To learn more about interactions between matter and dark matter on the 
microscopical level
the general covariant Dirac equation is derived in the Minkowski space--time
and in course of this the fundamental role of
internal symmetry is recognized. 
On this ground the Dirac equation are derived which
describe the interactions of spinor field with dark matter field.      
From this it follows the general conclusion that interactions of generalized
electromagnetic field with Dirac spinor field occur only via the 
electromagnetic field and the above-introduced dark matter field vacuum. 
The general 
conclusion is that
a dark matter gravitate but there is no actually  direct
interactions of this new form of matter  with known 
physical fields that represent luminous matter.
A rather simple and
feasible experiment is proposed to verify  this conclusion. 
The paper  is organized as follows. The first two sections are the basis for 
all  considerations.
Section 2 contains the geometrically motivated general definition of the  
concept of potential field. 
The conjecture is put forward that all potential fields has a
geometrical interpretation.
It is shown that in general case a parallel transport  is a exact
realization of the abstract concept of potential field. We consider this 
realization as a new physical field (dark matter field).
In Section 3  the equations of the dark matter field are derived.
Section 4 deals with the vacuum of this field. The equations of vacuum  field
are considered in Section 5 and 6. 
Section 7 treats the general covariant
Dirac equation in the Minkowski space--time  with careful
consideration of internal and space--time symmetries and connection between
them. In Section 8 the theory of interactions of the mentioned above
potential fields with matter  (spinor field) is formulated. The source
of the dark matter vacuum field is the circulation of the energy of the 
spinning matter, which is expressed in the direct connection between the 
potential of the dark matter vacuum 
field and the canonical tensor energy-momentum of spinning matter.
Thus, it is shown that the canonical energy-momentum tensor plays fundamental
role in the theory of the spinor fields.
And finally, Section 9 provides
a  proposal of rather simple experiment that can give answer 
the series of principal questions.

 \section{ Concept of potential field }
First of all we shall consider the necessary elements of general mathematical 
structure. According to the modern viewpoint a fundamental physical theory is
the one that possesses a mathematical representation whose elements
are smooth manifold and geometrical objects defined on this manifold.
Most physicists nowadays consider a theory be fundamental only if it
does make explicit use of this concept. It is thought that curvature
of the manifold itself provides an explanation of gravity. Within
the manifold, further structures are defined including vector fields,
connexions, particle path and so forth, and these are taken into
account for the behavior of physical world. This picture is 
generally accepted and it is based on such a long history of physical
research, that there is no reason to question it. The another element is the
concept of potential field.

If we
take  the components of symmetrical covariant tensor field  $g_{ij}$ and 
form its derivatives
($\partial_i g_{jk}$ ) then these derivatives are neither the components of
a tensor nor of any geometrical object. However, from $g_{ij}$ and these partial
derivatives one can form (with help of algebraic operations only) a new 
geometrical object 
\begin{equation}
 {\Gamma}^i_{jk}  = \frac{1}{2} g^{il}(\partial_j g_{kl} + \partial_j
g_{kl} -\partial_l g_{jk} ) \label{1},\end{equation} 
which is called Christoffel
connection, where $g^{il}$ are
contravariant components of the $g_{ij}.$ Now we can formalize this particular
case and give general definition of the potential field. 
        
If some geometrical object (or a geometrical quantity)  is given and from
the components of this object and its partial derivatives one can form
(using the  algebraic operations only) a new geometrical object (or 
geometrical quantity), then we deal with a new geometrical quantity that 
will be called a potential
field. Potential field is characterized by the
potential $P$ and strength $H$ and in what follows will be written in the form
(P,H). Connection between the potential and strength  is
then called  a natural derivative and in symbolic form can be written as
follows $H = \partial P.$ If we go back to our starting point, then
$g_{ij}$ is a potential and ${\Gamma}^i_{jk}$ 
is a strength of potential field 
$(g, {\Gamma} )$, known after Einstein as the gravitational field. 

Now we introduce another very important and geometrically motivated potential
field. The most important  geometrical notions are the
metric $g_{ij}$ and parallel transport or linear (affine) connexion $P^i_{jk}.$
Tensor field $g_{ij}$ is symmetric, $g_{ij} =
g_{ji},$  but linear connexion $P^i_{jk} $ in general is nonsymmetric
with respect to the covariant indices, $ P^i_{jk} \neq
P^i_{kj} $  and in any way does not link
with the metric $g_{ij}.$ In fact, these notions define, on a
manifold $M$, different geometric operations. Namely, a metric on a
manifold defines at every point the scalar product of vectors from
the tangent space and linear connection gives the parallel transport 
of these vectors along any path on M.
Consider a vector field $E^i(x).$  Equation of local parallel transport
from a point $x^i$ to a point $ x^i+dx^i $  has in general the form
 \be  dE^i(x) = - P^i_{jk}(x) E^k(x) dx^j \label{2} ,\ee 
where  functions $P^i_{jk}(x) $ are
components of a new geometrical object on the manifold, called a  
linear connexion $P$.  Under a parallel transport along the
infinitesimal closed curve the change  of the vector is equal to the quantity
$$     \triangle E^k = - H_{ijl}{^k} E^l dx^i \delta x^j,  $$
where
\be
   H_{ijl}{^k} = \partial_i P^k_{jl} -  \partial_j P^k_{il} +
P^k_{im} P^m_{jl} - P^k_{jm} P^m_{il} \label{3}
\ee
is a tensor field of the type (1,3), called the Riemann tensor of
the  connection $ P^i_{jk} .$
Now we go back to the definition of potential field and see that parallel 
transport defines new potential field $(P, H).$ At first glance
this is in contradiction with fundamental principle, which means that only 
irreducible quantity should enter into the theory. Indeed,
from (2) it follows that under a coordinate mapping
 $$\tilde x^i = \tilde x^i(x), \quad x^i = x^i(\tilde
x),$$ the transformation law for a  $ P^i_{jk} $    has the form
 \begin{equation} \tilde P^i_{jk} = \frac{\partial
\tilde x^i}{\partial x^l} ( P^l_{mn} \frac{\partial x^m}{\partial \tilde x^j}
\frac{\partial x^n}{\partial \tilde x^k} + \frac{\partial^2 x^l}{\partial
\bar x^j \partial \tilde x^k } ) \label{4}.  \end{equation}
Recall that a geometrical quantity is reducible if it is possible
to find linear combinations of its components which themselves constitute a
new geometrical quantity. As for linear connection  under the coordinate
mappings it is a reducible quantity which  is easily seen
from the expansion $$ P^i_{jk} =\frac{1}{2} (P^i_{jk} + P^i_{kj}) +
\frac{1}{2} (P^i_{jk} - P^i_{kj}).$$ From (4) it follows that a symmetrical
part of the connection $  \frac{1}{2}
(P^i_{jk} + P^i_{kj}) ,$ is again the linear connection and the
antisymmetrical part, $ \frac{1}{2}
(P^i_{jk} - P^i_{kj}) $ transforms  as a
tensor field of  the type (1,2).  However, there is a very interesting 
structure which allows to consider parallel transport as potential field.

Let $S^i_j$ be components of a tensor field of the type $(1,1)$
(a field of linear operator),
 ${\rm Det} (S^i_j) \neq 0.$ Out of two tensor fields $S^i_j$ and
$Q^i_j$ of the type (1,1) a tensor field $P^i_j = S^i_{k} \,Q^k_j $
of the type (1,1) may be constructed, called their product. With the
operation of multiplication thus defined, the set of tensor fields of
the type (1,1) with a nonzero determinant forms a group, denoted by
$G_i.$ This is a natural group of local symmetry on a manifold.
At given vector field $E^i,$  any element of the group $G_i$  
defines a bundle of vector fields, which is defined as follows
$$ {\bar E}^i =S^i_j E^j , \quad {\tilde E}^i = T^i_j E^j, \quad \mbox{etc},$$ 
where $T^i_j$ are components of the field $S^{-1}$ inverse to 
$S,\quad S^i_k T^k_j = \delta^i_j.$ It is 
clear that notion of the parallel 
transport is not applied to the bundle of the vector fields. From (2) it
follows that the parallel transport of the bundle of the vector fields is 
defined by the bundle of the linear connections, which is defined by 
the relation
$$ 
\bar P^i_{jk} =S^i_m P^m_{jn} T^n_k + S^i_m  \partial_j T^m_k.$$ 
It is easy to see  that for the bundle of linear connections the 
expansion  considered
above has no sense, so the tensor $P^i_{jk} - P^i_{kj} $ is evidently not
a geometrical quantity with respect to the transformations of the local group. 

Thus, we shall expand the diffeomorphism group to include 
into  the consideration the group of local symmetry $G_i,$ defined above. 
It can be shown that the diffeomorphism group is the group of
external automorphisms of the  group of local symmetry, i.e.  the
group $G_i$ is invariant under the transformations of the group ${\rm
Diff}(M).$ Thus, we have a nontrivial unification of these
symmetries and possibility to consider one more potential field. 

We conclude that we  really introduce geometrically motivated
potential field $(P, H)$, but the theory of this field should be invariant  
not only with respect to the general transformations of the coordinates but 
with respect to the transformations of the  local symmetry group $G_i$ as well.
We put in correspondence to this field  the so called dark matter and develop 
theory of the dark matter as the theory of this new potential field.

For brevity, we will use in what follows the matrix notation
$$ P_j = (P^i_{jk}), \quad E= (\delta^i_j), \quad H_{ij} =(H_{ijl}{}^k),
\quad S= (S^i_j), \quad {\rm Tr}S = S^i_i, $$  in which the transformation law
of the potential $P_j$ is of the form
\be
\bar P_{j} =S P_{j} S^{-1} + S\partial_j S^{-1} = P_i + SD_i S^{-1},\label{5}
\ee
where $D_i$ stands for the important operator
     $$D_i S = \partial_i S + P_i S - S P_i =\partial_i S + [P_i, S],$$
which is especially convenient when one deals with local symmetry in 
question. In what follows we shall meet many examples of this.
The relation (5) is indeed the transformation of the connection, since
$SD_i S^{-1} $ is a tensor field of the type (1,2) and on this reason 
$\bar P_{j} $ is the connection with respect to the coordinate 
transformations. Since the connection between the potential and strength 
in matrix notation is given by the formula
$$  H_{ij} = \partial_i P_j - \partial_j P_i + [P_i, P_j],$$
from (5) it follows that under the transformations of the group $G_i$
the strength is transformed as follows 
\begin{equation} \bar H_{ij}  = S H_{ij} S^{-1}. \label{6}
\end{equation} 
For the $H_{ij}$  we have
$$ D_i H_{jk} = \partial_i H_{jk} + [P_i , H_{jk}]$$
and if $\bar D_i$ is defined by potential $\bar P_i,$ then from (5) and (6) 
it follows  that
$$\bar D_i \bar H_{jk} = S (D_i H_{jk}) S^{-1}. $$
In  general case the operator $D_i$ is not general covariant,
however, the commutator $[D_i , D_j]$ is always general covariant and we get 
the important relation for the strength tensor of dark matter  
\begin{equation} [D_i , D_j ] H_{kl} = [H_{ij} , H_{kl}].  \label{7}
\end{equation}

Thus, in our approach the theory of the dark matter is tightly connected with
the local symmetry, it is  general covariant and has a profound geometrical 
interpretation.

 \section{ Field Equations }

The simplest general covariant and gauge invariant Lagrangian of the 
potential $P_i$ is a direct consequence of  (6) 
\begin{equation}
    L_{P} = - \frac{1}{4}{\rm Tr} (H_{ij}  H^{ij}),   \label{8}
\end{equation}
where  $ H^{ij}= g^{ik}g^{jl} H_{kl}.$  
Varying the Lagrangian $ L_{P}$ with respect to $P_i$  and using the
relation $\delta H_{ij} = D_i {\delta} P_j - D_j {\delta} P_i$ we
obtain 
$$\delta L_{P} = {\rm Tr}((\frac{1}{\sqrt{g} } D_i (\sqrt {g} H^{ij})) 
\delta P_j)  - \frac{1}{\sqrt{g}} \partial_i {\rm Tr}(\sqrt {g} H^{ij} 
\delta P_j ) $$ and  hence 
the following equations of the  field hold valid 
\begin{equation}
 \frac{1}{\sqrt{g} } D_i (\sqrt {g} H^{ij}) = 0, \label{9}         
\end{equation}
where $ g = -{\rm Det}(g_{ij}). $ 
From the properties of the operator $D_i$ it is not difficult to see that 
equations (9) are invariant with respect to the local symmetry group.
The tensor character of these equations can be seen from the identity
$$\frac{1}{\sqrt{g} } D_i (\sqrt {g} H^{ij})= \stackrel {p} {\nabla}_                        i H^{ij}
+ \omega_i H^{ij} - \frac{1}{2} (P^j_{ik} - P^j_{ki}) H^{ik},$$
where $\stackrel {p} {\nabla}_i$ is the usual covariant derivative with 
respect to the connection $P_i$ and 
$\omega_i= \partial_i\ln\sqrt{g} -P^k_{ki} $ are the components of the
covector field. Thus, it is shown that the group of diffeomorphisms is the 
group of covariance of the equations (9).
The equations (9) form first group of the equations. The second one is 
presented by the identity
   \be D_i H_{jk}+ D_j H_{ki}+ D_k H_{ij} =0. \label{10}
   \ee
From definition of the operator $D_i$ it follows that left hand side of the
relation (10) is a tensor and hence it is general covariant.

Varying the Lagrangian $ L_{P}$  with respect
to $g^{ij}$ we obtain the so--called metric tensor of energy--momentum of the
dark matter field \begin{equation} T_{ij} = 
   {\rm Tr} (H_{ik} H_j{}^k) + g_{ij} L_{P},\label{11} \end{equation} where $
H_j{}^k = H_{jl} g^{kl}.$ One can establish the identity
$$\nabla^i T_{ij} = {\rm Tr}(H_{jk}\frac{1}{\sqrt{g} } D_i (\sqrt {g} H^{ik})
) + \frac{1}{2} {\rm Tr} (H^{ik}(D_i H_{jk}+ D_j H_{ki}+ D_k H_{ij} )). $$
With this and equations (9) and (10) we see that the metric tensor of the
energy--momentum satisfies the  equations
\be \nabla_i T^{ij} = 0, \label{12} \ee
where the $\nabla_i $ denotes as usual the covariant derivative  with respect to
the Christoffel connexion (1) and $\nabla^i = g^{ik} \nabla_k.$ It is 
evident that the metric tensor energy--momentum is invariant with respect 
to the group of local transformations in question.  
Now we can write down the full action for the fields
$g_{ij}$ and $ P_i $
$$ A = -
\frac{c^3}{G}\int R \sqrt{g} \, d^4 x -\frac{\beta^2 \hbar}{4}\int {\rm Tr}
(H_{ij} H^{ij}) \sqrt{g} \, d^4 x, 
$$
where $R$ is the scalar curvature, $G$
is the Newton  gravitational constant, $\hbar$ is the Planck constant and
$\beta $ is dimensionless constant.
From the geometrical interpretation of the field $P$ it
follows that it has the dimension $cm^{-1}.$
As all coordinates can be considered to have the dimension $cm,$  the action
$A$ has a correct dimension. 

Varying the full action $A$ with respect to $g^{ij}$ we derive the
Einstein equations
\be R_{ij} - \frac{1}{2} g_{ij} R = \beta^2 l^2 T_{ij}, \label{13} \ee
where $l = \sqrt{\hbar G/c^3}$  is the Planck length and $T_{ij}$  is the
metric tensor of energy-momentum of dark matter field. Thus, it is shown 
that the interactions of the dark matter field with the gravitational field 
are characterized by some length $\lambda = \beta l.$  Equations (9),(10) 
and (13) are compatible in view of (12).

Equations (9) and (10) constitute the full system of the generalized Maxwell
equations in geometrical representation and new field (dark matter field) can 
be considered as the generalized electromagnetic field. The arguments are as
follows.  From the strength of the potential field in question it can be 
constructed very interesting quantity that is invariant with respect  to the
transformations of the group of  local symmetry $G_i,$ namely
\be
   F_{ij} = {\rm Tr} H_{ij}.    \label{14}
\ee
It is evident that $F_{ij} $ is an antisymmetrical tensor with respect to the
transformations of coordinates.
If $H_{ij}$  satisfy the equations (9) and (10) then taking the trace we 
obtain that bivector $F_{ij} $ satisfies the Maxwell equations
\be
     \frac{1}{\sqrt{g} } \partial_i (\sqrt{g} F^{ij}) =0, \quad 
  \partial_i  F_{jk} + \partial_j F_{ki} + \partial_k F_{ij} =0. \label{15}
\ee
Now consider a question concerning the vector potential of the electromagnetic 
field. We put $A_i = {\rm Tr} P_i = P^k_{ik}.$  According to (5) and the 
differentiation rule for determinants, the transformation law for $A_i$ under 
the local transformations has the form
$$ \bar A_i = A_i - \partial_i ln|\bigtriangleup|, $$ where 
$\bigtriangleup ={\rm Det}(S^i_j).$  Thus, the local transformations of 
potential $P^i_{jk}$ reduced to the gauge transformations of the potential of 
the electromagnetic field $A_i.$  For completeness of the picture we shall also 
consider the arbitrary coordinate transformations of $A_i.$  From (4) one 
can derive that $A_i$ transforms as follows
$$\tilde A_i = (A_m - \partial_m ln|J|) \frac{\partial x^m}{\partial \tilde x^i},
$$ where $J = |\frac{\partial \tilde x}{\partial x}|$ is the Jacobian
of the transformation. It is interesting to point out that any arbitrary 
coordinate transformation is accompanied by the gauge transformation.
Since $F_{ij}= \partial_i A_j - \partial_j A_i, $ the question on the nature
of the gauge transformations is completely solved and geometrical origin of 
the electromagnetic field is recognized. 

Now we have to solve two problems.  If generalized electromagnetic field 
represents dark matter it should be massive (whereas electromagnetic field is 
massless) and the  other problem is the general covariant gauge fixing that is 
provided by the 
Cauchy problem for the field in question. The distinctive feature of the 
generalized electromagnetic field is that it is self--interacting: it is 
non-linear even in the absence of other fields. Two potentials $\bar P_i$ and 
$P_i$ are physically equivalent if there is a local transformation which takes
$P_i$  into $\bar P_i,$ and clearly $\bar P_i $ satisfies the field equations if 
and only if $P_i$  does. In order to obtain a definite member of the equivalence
class of potentials one has to introduce  general covariant
gauge conditions. These conditions have to remove the sixteen degrees of 
freedom and lead to unique solution for the 
potential components.  To solve these problems we suggest that gauge fixing 
is an internal property of the system in question and introduce very important
notion of the vacuum of generalized electromagnetic field.

\section{Vacuum}
We have a vacuum if $H_{ij} =0$ and so the energy density of the generalized 
electromagnetic field is equal to zero. 
On the other hand, $P_i \neq 0$  so the 
vacuum has a structure. 
Let four linear independent covector fields be given
$E_i^{\mu}, \quad p={\rm Det}(E_i^{\mu} ) \neq 0 .$  Greece indices belong 
to the internal symmetry which
we shall in what follows connect with internal symmetry inherent in the
Dirac equation, whereas latin indices are coordinate.
Under a general transformation $ x^i = {\tilde
x}^i(x)$ of the coordinate system, each of these fields transforms as
follows $$ \tilde E_i^{\mu} = E_k^{\mu} \frac{\partial { x}^k}{\partial \tilde
x^i}, \quad \mu=0,1,2,3. $$ From the $E_i^{\mu}$ one
can purely algebraically construct components of the four vector field
$E_{\mu}^i$ so that \be E^i_{\mu}
E^{\mu}_j = \delta ^i_j, \quad E^i_{\mu} E^{\nu}_i = \delta^{\nu}_{\mu}
\label{16}\ee hold valid. If we  put $P^i_{jk} =V^i_{jk}, $  where
\be
V^i_{jk} = E^i_{\mu} \partial_j E^{\mu}_k, \label{17}  \ee then it is easy to
show that this is  the solution of the vacuum equation
$H_{ij} = 0$ for any $E^{\mu}_k. $ If $\bar V_i$  is another solution then it 
can be shown that
$\bar E^i_{\mu} = S^i_k E^k_{\mu} $ and $\bar V_i = S V_i S^{-1} + 
S \partial_i S^{-1}.$  Thus, the vacuum of generalized electromagnetic field
is again a potential field (E, V) with $E_i^{\mu} $  being potential and $V_i$ 
being strength. 

Now we introduce the tensor field
\be
   Q^i_{jk} = P^i_{jk}- V^i_{jk},  \label{18}
\ee
which can be called the deviation of the generalized electromagnetic field 
with respect to a vacuum. It is evident that under the local transformations
the deviation tensor transforms as follows
$$ \bar Q_i = S Q_i S^{-1}. $$
The tensor $Q$ is reducible and in what follows we shall consider the 
irreducible deviation tensor
$$ T^i_{jk} = Q^i_{jk} - \frac{1}{4} Q^m_{jm} \delta^i_k, \quad T_j = Q_j 
-\frac{1}{4} ({\rm Tr}Q_j) E.$$  With this we can consider the general 
covariant and gauge invariant Lagrangian
of the generalized electromagnetic field in the following form
\begin{equation}
 L_{P} = - \frac{1}{4} {\rm Tr} (H_{ij}  H^{ij}) - \frac{\mu^2}{2}{\rm Tr} 
 (T_i T^i),
 \label{19}
\end{equation}
where $\mu$  is a constant, which has dimension $cm^{-1}.$  It is natural to 
identify this constant with the length that characterizes the interactions
of the dark matter field with gravitational field, $\mu = 1/ \lambda.$
Varying  (19) with respect to $P_i$ we get the following equations
\begin{equation}
 \frac{1}{\sqrt{g} } D_i (\sqrt {g} H^{ij}) = \mu^2 T^j.\label{20}         
\end{equation}
We see that in some sense one can treat $\mu$ as the effective mass of 
the heavy photon. Since trace of $T^i$ equals zero, 
from (20) it follows that photon
remains massless. From (20) it follows that $T^i$ has to satisfy the equation
\begin{equation}
 \frac{1}{\sqrt{g} } D_i (\sqrt {g} T^{i}) = 0,   \label{21}       
\end{equation}
since in accordance with (7)  $ D_iD_j (\sqrt {g} H^{ij})=0. $
It is very important that the same equation appears under varying (19)  
with respect to $E^i_{\mu}.$ The equation (21) represents sixteen additional
constraints on the potential $P_i.$

However equations (20) and (21) are invariant with respect to the local 
transformations and hence we still have problem of gauge fixing.
To find its natural solution we can look for the geometrically motivated 
equations for the vacuum field, which are not invariant with respect to the 
transformations  of the local symmetry group $G_i.$  It is interesting that 
such possibility really exists.

\section{Equations of the vacuum field}
The local symmetry will be broken if we introduce the quantity
 \be U^i_{jk} = E^i_{\mu}
(\partial_j E^{\mu}_k - \partial_k E^{\mu}_j). \label{22} \ee
From the definition it follows that $U^i_{jk}$ is evidently a tensor field 
antisymmetric in covariant indexes. On the other hand, from (5)  it follows 
that this tensor is not geometrical object with respect to the local symmetry 
group.  The tensor $U^i_{jk}$ defines no representation of the group $G_i.$
Thus, it is convenient for our goal.  Further we shall establish geometrically 
motivated Lagrangian  that can be constructed for this vacuum tensor field. 
It leads us to the investigation of the geometry of 
affine  space which is characterized by the connection 
\be L^i_{jk} =\Gamma^i_{jk} + U^i_{jk}
\label{23},\ee 
where first summand is given by the expression (1).
Physical meaning of this connection is to investigate two quite independent
potential fields in the uniform geometrical framework.
Consider the most important geometrical quantity defined by the connection
(23). For the Riemann tensor  as a function of the potentials of
gravity and vacuum we have
\be  B_{ijk}{^l} = R_{ijk}{^l} + \nabla_i U^l_{jk} - \nabla_j U^l_{ik} +
U^l_{im} U^m_{jk} - U^l_{jm} U^m_{ik} \label{24}, \ee 
where \be R_{ijk}{^l}
= \partial_i \Gamma^l_{jk} - \partial_j \Gamma^l_{ik} + \Gamma^l_{im}
\Gamma^m_{jk} - \Gamma^l_{jm} \Gamma^m_{ik} \label{25} \ee 
is the Riemann curvature tensor of metric $g_{ij}$ and $\nabla_i$ as earlier
stands for 
the covariant derivative with respect to the Christoffel connection (1) 
$$\nabla_i
U^l_{jk} = \partial_i U^l_{jk} + \Gamma^l_{im} U^m_{jk} -\Gamma^m_{ij}
U^l_{mk} -\Gamma^m_{ik} U^l_{jm}.$$ By contraction we get from (24) the
tensor \be B_{jk} = B_{ijk}{^i} =
R_{jk} + \nabla_i U^i_{jk} - \nabla_j U^i_{ik} + U^i_{im} U^m_{jk} -
U^i_{jm} U^m_{ik} \label{26}, \ee 
where $R_{jk}$ is the Ricci tensor. From  (26) one can find by
contraction with metric the following expression for the scalar:
  $$ B = g^{jk}B_{jk}=
R + g^{jk} U^l_{jm} U^m_{kl} - \nabla_j U^j, $$ where $R$ is the Ricci
scalar curvature  and $U^j = g^{jk} U_{k} = g^{jk} U^l_{lk}.$ Hence, 
connection (23) uniquely determines the geometrical Lagrangian of the 
potential
fields of curvature and vacuum which is a natural generalization of
the Einstein--Gilbert Lagrangian of the gravitational field.
Thus, we shall derive equations describing the interactions of the
gravitational and vacuum fields from the action
 \be A =\frac{c^3}{2G} \int{B \sqrt g} d^{4}x .\label{27} \ee From (27) it
follows that connection (23) uniquely determines the Lagrangian  $L_v$ of
the vacuum field itself    \be L_v =
\frac{1}{2} g^{jk} U^l_{jm} U^m_{kl} \label{28}.  \ee It is natural
that the Lagrangian of the vacuum field like the dark matter
Lagrangian contains no  derivatives of the components of the
gravitational potential, since $U^i_{jk}$ can be considered as a strength with 
respect to $E^{\mu}_i.$

To conclude this section, we establish one more interesting connection
between two potential fields in question. Standard Lagrangian of the
gravitational field  $L_g = R $ contains a second order
derivatives of $g_{ij} $ and this leads to the known difficulties
~\cite{4}. Let us  show, that this Lagrangian can be generally
covariantly reduced to the Lagrangian without a second order
derivatives of $g_{ij}. $

Introduce a binary tensor field
\be B^i_{jk} =   E^i_{\mu} \nabla_j E^{\mu}_k = V^i_{jk}
-{\Gamma}^i_{jk} \label{29}.  \ee 
Setting $$
V^i_{jk}=\Gamma^i_{jk} + V^i_{jk} -\Gamma^i_{jk} = \Gamma^i_{jk} +B^i_{jk},
$$ and following closely the line defined by (24) and (26), we derive the
relation $$ 0 = R_{jk} + \nabla_i B^i_{jk} - \nabla_j B^i_{ik} + B^i_{im}
B^m_{jk}- B^i_{jm} B^m_{ik}.$$ From the last formula it follows that $$ R +
\nabla_i(g^{jk} B^i_{jk} - g^{ik} B^l_{lk})= g^{jk} ( B^i_{jm} B^m_{ik} -
B^i_{im} B^m_{jk}).$$ Thus, the Einstein--Hilbert Lagrangian is
equivalent to the Lagrangian
$$L_{gv} = g^{jk} ( B^i_{jm} B^m_{ik}
- B^i_{im} B^m_{jk}) ,$$ which is defined by the vacuum field and may be
more convenient in the quantum theory of the gravitational field.

\section{Curvature and vacuum in interaction}
Varying action (27) with respect to  $g_{ij},$ we get the Einstein
equations  $$ G_{ij} = T_{ij} ,$$ where 
\be
T_{ij} = g_{ij} L_v - U^k_{il} U^l_{jk} \label{31} \ee 
is the metric tensor
energy--momentum of the vacuum field.  From (28) and (30) it follows that
$g^{ij} T_{ij} = 2 L_v $ and hence equations of the vacuum field are
not conformally invariant.  It is yet another general property of gravity and
vacuum fields.

Now we make small variations in our field quantities $E^l_{\mu}.$ It
is convenient to introduce tensor
$$F^{ij}_k =g^{il} U^j_{lk} - g^{jl} U^i_{lk} = U^{ij}_k - U^{ji}_k $$
with inverse transformation
$$U^i_{jk} = \frac{1}{2}(g^{il} F^{mn}_l g_{jm} g_{kn}  + g_{jl} F^{il}_k -
g_{kl} F^{il}_j ). $$  Since $$ U^i_{jk} = E^i_{\mu} (\partial_j E^{\mu}_k
- \partial_k E^{\mu}_j )= E^i_{\mu} (\nabla_j E^{\mu}_k - \nabla_k
E^{\mu}_j ) , $$ we get sequentially (28), 
\be \delta B= \delta L_v =
 F^{jk}_l (\nabla_j
E^{\mu}_k) \delta E^l_{\mu} + F^{jk}_l E^l_{\mu} \nabla_j \delta E^{\mu}_l
\label{31}  .  \ee  
With (16)  $$ \delta E^{\nu}_k = - E^{\nu}_l
 E^{\mu}_k \delta E^l_{\mu}. $$ By this, the second term in the right hand
 side of  (31) can be presented in the following form
  $$ \nabla_j (F^{jk}_l E^l_{\mu} \delta E^{\mu}_l) +  E^{\mu}_k
 (\nabla_j F^{jk}_l + F^{jk}_m E^{\nu}_l \nabla_j E^m_{\nu} ) \delta
 E^l_{\mu} .$$ Thus, the variational principle provides the following
 equations for the potential of the vacuum field $$ E^{\mu}_k
 \nabla_j F^{jk}_l  + F^{jk}_l \nabla_j E^{\mu}_k + F^{jk}_m E^{\mu}_k
 E^{\nu}_l \nabla_j E^m_{\nu} = 0 .$$ It is possible to rewrite
 this equations in more symmetrical form (without covariant
 derivative of the potential).
 With (16) and (29) we have $$ E^{\nu}_l \nabla_j
 E^m_{\nu}= -E^m_{\nu} \nabla_j E^{\nu}_l= - B^m_{jl}, \quad \nabla_j
 E^{\mu}_k= B^m_{jk} E^{\mu}_m $$ and hence equations of the vacuum
 field have the following form
  \be \nabla_j F^{jk}_l + B^k_{jm} F^{jm}_l - B^m_{jl} F^{jk}_m = 0
 \label{32}.  \ee Like the equations of the gravitational field and 
 dark matter field the equations of the vacuum field are essentially nonlinear. 
 Let $\stackrel{v} {\nabla}_i$ be a covariant derivative with
 respect to the  connection (17).  Since $$\nabla_j F^{jk}_l =
 \stackrel{v} {\nabla}_j F^{jk}_l -B^j_{ji} F^{ik}_l -B^k_{jm}
 F^{jm}_l + B^m_{jl} F^{jk}_m,$$ equations of the vacuum field (32) can be
 presented in the following most simple form
 \be (\stackrel{v} {\nabla}_j-B_j) F^{jk}_l  =0 \label{33}, \ee
 where $B_i$ is a contraction of the binary tensor field (28), $B_i = B^k_{ki}.$

In conclusion of this section 
we would like to point out on possible applications of the
equations derived. 
It is of interest to find spherically symmetric solution of
the system of equations (Einstein  equations plus (33)) and then 
investigate the corresponding metric of the generalized Schwarzschild
solution. 

Now we shall consider the interactions of generalized 
electromagnetic field with matter in the framework of the Dirac theory that
is very important since it is known nothing about the interactions of the dark
matter field with luminous matter.

\section{The Dirac equation in general covariant form}

The description of the interactions between the matter and dark matter we will
provide in the framework  of the Dirac equation, which is the basis for the
description of matter. 
It is one of the fundamental principles of modern geometry and theoretical
physics that laws of geometry and physics do not depend on the choice  of
coordinate systems. It is natural to write all equations in the coordinate
basis since the problem to rewrite these equations in any other basis is
formal and hence trivial task. In our days this statement is as canonical
as the energy conservation.  Let us show that original Dirac equation is in 
full agreement with this fundamental statement and that it is defined by the 
internal symmetry.
As it is known, internal
symmetries play fundamental role in modern physical theories and hence it
is very important to have clear understanding of  the role of internal
symmetries in the Dirac equation, which is the basis for all modern
theories of elementary particles and their interactions, in particularly,
Dirac's Hamiltonian defines entirely the space--time sector of the standard
model.

Let ${\bf C}^4 $ be a linear space of columns of four complex
numbers $\psi_{1}, \,\psi_{2},\, \psi_{3},
\,\psi_{4}.$ Linear transformations in this space can be presented by
the  complex matrices
$(4 \times 4).$ The set of all invertible $(4 \times 4)$ complex matrices
forms a group denoted by $GL(4,{\bf C}).$ Dirac's  $\gamma^{\mu}$ matrices
belong to $GL(4,{\bf C}) $ and obey anticommutation relations  $$
\gamma^{\mu}\gamma^{\nu}+ \gamma^{\nu}\gamma^{\mu} = 2 \eta^{\mu \nu},$$
where  $\eta^{\mu \nu}$ is digital matrix such as the inverse
matrix $\eta_{\mu \nu}$ defines the commutation relations of the Poincar\' e 
group. In the case of the Poincar\' e group it is possible to write the 
structure
relations with help of matrix $\eta_{\mu \nu}$ and signs plus and
minus but for our consideration the explicit form of the matrix 
$\eta_{\mu \nu}$ is not
important. One should only not confuse the $\eta_{\mu \nu}$
with the Minkowski metric $g_{ij}$, which has quite another sense.

From $\gamma^{\mu}$ one can
construct sixteen linear independent matrices that form a basis of the Lie
algebra of $GL(4,{\bf C}).$ This basis is especially important since the
matrices $S_{\mu \nu} = \frac{1}{4}( \gamma_{\mu}\gamma_{\nu}-
\gamma_{\nu}\gamma_{\mu} ) $ form the
basis of the Lie algebra of the Lorentz group (subgroup of $GL(4,{\bf
C}).$) Thus, we suppose that the Dirac spinor is an element of the space 
${\bf C}^4 $ where the group $GL(4,{\bf C})$ acts that is
equipped with the matrix $\eta_{\mu \nu}.$ For better understanding it should
be noted that in the space ${\bf C}^3 $ there are no matrices like
$\gamma^{\mu}.$

If one considers $\psi_{1}, \,\psi_{2},\, \psi_{3}, \,\psi_{4}$
as a set of complex scalar fields on the space--time manifold then
the Dirac spinor field emerges on the manifold which is a basis of irreducible
representation of the group $GL(4,{\bf C}).$  It is not difficult to
understand that $GL(4,{\bf C}) $  is a group of internal symmetry since
its transformations involve only functions of the spinor field and do not
affect the  coordinates. In other words, spin symmetry is {\bf an internal
symmetry}.

Now, on this ground we consider general covariant formulation of the Dirac
equation in the Minkowski space--time. We shall follow the fundamental
physical principle that was mentioned above. With respect to an
arbitrary curviliner system of coordinates Minkowski space--time is
characterized by the metric $$ ds^2 = g_{ij} dx^i dx^j$$ of the Lorentz
signature, which satisfies the equation $R_{ijk}{^l} =0 $ and topology $R^4.$
At given $g_{ij},$
the generators of the group of space--time symmetry can be presented as a
set of linear independent solutions of general covariant system of
equations (Killing's equations)$$ K^i \partial_i g_{jk} + g_{ik} \partial_j
K^i + g_{ji} \partial_k K^i =0 $$ for a vector field $K^i.$  In the case of
the Minkowski metric we have ten linear independent solutions of the
Killing equations, which are denoted $K^i_{\mu}$  and $K^i_{\mu \nu} = -
K^i_{\nu \mu} $ and  hence the Greek indices  enumerate  vector
fields and take the values $0,1,2,3,$ like  coordinate Latin indices.

It is well--known that the generators of the Poincar\'e group
$$ {\bf P_{\mu}} = K^i_{\mu} \frac{\partial}{\partial x^i}, \quad
  {\bf M}_{\mu \nu} = K^i_{\mu \nu} \frac{\partial}{\partial x^i} $$
satisfy the following commutation relations
\begin{equation} [{\bf{P}}_{\mu}, {\bf{P}}_{\nu}] =0, \label{34}
\end{equation}
\begin{equation} [{\bf{P}}_{\mu}, {\bf{M}}_{{\nu}{\lambda}}]
 = \eta_{\mu \nu} {\bf{P}}_{\lambda} - \eta_{\mu \lambda} {\bf{P}}_{\nu}.
 \label{35} \end{equation}
It is evident that  all these relations are
general covariant and that the operators
  ${\bf P}_{\mu}=  K^i_{\mu} \frac{\partial}{\partial x^i} $
transform a scalar field into the scalar one.

Now we shall show that the general covariant Dirac equation has the form
 \begin{equation}
            i\gamma^{\mu} {\bf{P}}_{\mu} \psi = \frac{mc}{\hbar }
 \psi, \label{36}   \end{equation}
 where  $\psi$ is a column of four complex scalar fields in
question and ${\bf {P}}_{\mu} $ are the generators of
space--time translations. To be exact in all details let us explain what 
does it mean that the Dirac equation
is general covariant. Transformation $\varphi$ of the local group of
diffeomorphisms (group of general coordinate transformations) can be
represented by the smooth functions
 $$\varphi: x^i \Rightarrow \varphi^{i}(x), \quad \varphi^{-1}: x^i
\Rightarrow f^{i}(x), \quad \varphi^{i}(f(x)) = x^i. $$
Induced transformation of the metric tensor is of the form
$$ {\tilde g}_{ij}(x) = g_{kl}(f(x)) f^k_i(x) f^l_j(x),$$ where $f^k_i(x)=
\partial_i f^k(x).$  For  the scalar and
vector fields we have
$$ {\tilde \psi}(x) = \psi(f(x)), \quad {\tilde P}^i(x) = P^k(f(x))
\varphi^i_k (f(x)), $$ where $\varphi^i_k (x) = \partial_k \varphi^i (x).$
It is not difficult to verify that if $K^i(x)$ is a solution of the Killing
equations for the metric $g_{ij} (x),$ then ${\tilde K}^i(x) $ is a
solution of the Killing equations for the metric ${\tilde g}_{ij}(x) .$
Further, if $\psi(x) $  is a solution of the Dirac equation (36), then
${\tilde \psi}(x) $ will be a solution of the equation (36) when $K^i_{\mu}(x)$
is substituted by the ${\tilde K}^i_{\mu}(x) .$  Besides
the transformations of the diffeomorphisms group conserve the form of the
commutation relations of the Poincar\'e group.
Dirac's equation is covariant with respect to the general coordinate
transformations.  It is known that in the Minkowski space--time there is
preferred class of the coordinate systems.  In the preferred system of
coordinates the Dirac equation (36) has a customary form.

It is also clear that the equation (36) is equivalent to
the equation $$ i{\tilde {\gamma}}^{\mu} {\bf {P}}_{\mu} \psi =
\frac{mc}{\hbar } \psi,$$  if $ {\tilde {\gamma}}^{\mu} = S \gamma^{\mu}
S^{-1},$ where $S \in  GL(4,{\bf C}) $ (the Dirac equation (36) is covariant 
with respect to the transformations of the group $ GL(4,{\bf C})). $ 

Now we have found enough to provide some valuable insights into the
connection between the space--time and internal transformations.
Consider again the generators of the internal Lorentz group
$S_{\mu \nu} = \frac{1}{4}( \gamma_{\mu} \gamma_{\nu} -
\gamma_{\nu} \gamma_{\mu}) $ and pay attention to the
commutation relations
\begin{equation} [\gamma_{\mu}, S_{{\nu}{\lambda}}] = \eta_{\mu \nu}
\gamma_{\lambda} - \eta_{\mu \lambda} \gamma_{\nu} \label{37}.
\end{equation} Comparing (35) and (37) it is not difficult to verify
that the operators  $${\bf{L}}_{\mu
\nu} = {\bf{M}}_{\mu \nu} + S_{\mu\nu} $$ commute with the Dirac
operator $D = i\gamma^{\mu} {\bf{P}}_{\mu} $  and satisfy  the commutation
relations of the Poincar\'e group. Thus, in the Minkowski space--time
there is a relation between the internal symmetry group and the space--time
symmetry group. The consequence is that Dirac's equation (36) is invariant
with respect to the transformations of the Poincar\'e group.
Thus, the geometrical and group--theoretical meaning of both spinor and
original Dirac equation is quite  clear.  We see that structure of the
Dirac equation is defined by the internal symmetry and the
derivatives with respect to the given directions. In considered case these
derivatives coincide with generators of the translation group.  In this
respect the Dirac equation   differs radically from the Einstein equation,
where internal symmetry have no role at all. The spinor enters into the
world of tensors as four component complex scalar field  being a carrier of
internal symmetry, which, thus,  was discovered together with the Dirac
equation.

Consider now the possible natural generalizations of the general
covariant Dirac equation.   We will strive to realize project
when diffeomorphisms group is the group of invariance (not covariance)  of
generalized theory and internal symmetry remains without change.
There is only one natural way to do this and it will be subject of our 
consideration in later sections.

\section{Generalization of the Dirac theory}

In this chapter it is shown that a spinor field can be presented as a
natural origin of the vacuum  potential field, considered above.

We take that  the canonical energy-momentum tensor
plays fundamental role in the theory of the spinor fields and in accordance
with this the generalized  Dirac's Lagrangian has the form
 \begin{equation} L_D =
\frac{i}{2}E^i_{\mu} \biggl(\bar \psi \gamma^{\mu} D_i \psi - (D_i \bar
\psi)\gamma^{\mu} \psi\biggr)- m \bar \psi \psi \label{38} , \end{equation}
where  $E^i_{\mu}$ are contravariant components of the potential of vacuum,
$$ D_i \psi = (\partial_i - iq A_i) \psi, \quad
D_i \bar \psi = (\partial_i + iq A_i) \bar \psi, \quad A_i = P^k_{ik}.$$  
It is evident
that varying  (38) with respect to $E^i_{\mu}$  results in canonical
energy-momentum tensor of the spinor field.
Lagrangian (38) is invariant with respect to the substitutions $$\psi
\Rightarrow e^{i\varphi} \psi, \quad \bar \psi \Rightarrow e^{-i\varphi}
\bar \psi, \quad A_i \Rightarrow A_i + \partial_i \varphi $$ and hence it is 
general covariant and invariant with respect to the local transformation of 
the group $G_i.$
Action has the form $$A = \hbar \int L_{D}
\, p \,d^4x,$$ where $p = \mbox{Det} (E^{\mu}_i).$ Since $$E^{i}_{\mu}
\partial_j E^{\mu}_i = \frac{1}{p} \partial_j p,$$ this action leads to
the Dirac equations in the presence of external vacuum and electromagnetic
fields \be i E^i_{\mu} \gamma^{\mu}
(D_i + \frac{1}{2} U_i) \psi = m \psi \label{39} ,\ee 
\be i E^i_{\mu}  (D_i
+ \frac{1}{2} U_i){\bar \psi} \gamma^{\mu} = -m {\bar \psi} \label{40} ,\ee
where, as earlier, $U_i= U^k_{ik}.$

Setting  $$
W^{\mu}_i = \frac{i}{2} (\bar \psi \gamma^{\mu} D_i \psi - (D_i \bar
\psi)\gamma^{\mu} \psi) , $$  we have $L_D =P^i_{\mu} W^{\mu}_i- m \bar \psi
\psi. $ Hence, from the action$$A =\hbar \int L_{D} \, p \,d^4x +
\frac{c^3}{G}\int L_v
\sqrt{g} d^4x, \quad g= -\mbox{Det}(g_{ij}) $$ we verify (in accordance
with (33))the following equations for the potential of the vacuum field
\be \nabla_j F^{jk}_l + B^k_{jm} F^{jm}_l - B^m_{jl} F^{jk}_m +l^2 W^k_l =0
\label{41}, \ee 
where $$W^k_l = \epsilon E^k_{\mu} W^{\mu}_l, \quad
\epsilon = p/ \sqrt g.  $$ The equations (41) generalize equations
(33) and together with the Dirac equations (39) and (40) explain
clearly how the vacuum field interacts with the spinor field.
The potential of the generalized electromagnetic field enter into the Dirac
Lagrangian only in the form of the trace of $P^i_{jk}.$  The other 
possibility does not exist.
From the equations (41) an interesting relation can be derived.
By summing over the indices  k and l we get that a trace of  $ U^i_{jk}$
satisfies the following equation \be \nabla_i U^i  = m {\bar \psi} \psi
\label{42}, \ee 
where $U^i = g^{ik} U_k.$ We conclude that for $m=0$ the
interactions of the vacuum and spinor fields are characterized by a new
conserved quantity. Indeed, this fact simply means that the action is
invariant under the mapping
$$E^{\mu}_i \rightarrow a E^{\mu}_i, \quad \psi \rightarrow
a^{-\frac{1}{2}} \psi,$$ where $a$ is dimensionless constant.
Thus, the introduction of the vacuum field into the framework of the
standard model may shed new light on the mechanism of the lepton mass
generation.

From the action
$$A =\beta^2 \hbar \int{L_{P}} \,\sqrt{g} \, d^4 x + \hbar \int{L_D} \,p \,d^4x $$
we derive the equations of the generalized electromagnetic field 
(dark matter field)
in interaction with the spinor field
\begin{equation}
 \frac{1}{\sqrt{|g|} } D_i (\sqrt {|g|} H^{ik}) + \frac{q}{\beta^2} J^k E = 
 \mu^2 T^j,\label{43}         
\end{equation}
where $$J^i = \epsilon E^i_{\mu} {\bar \psi}\gamma^{\mu} \psi $$  is the 
Dirac vector of current.   
Thus, the basic equations of the considered fields are derived.

Now it is important to exhibit the equations that follow from the equations
derived. To this end we shall establish the identities for the Lagrangians
of the fields in question.  
Identity for the Lagrangian of the vacuum field may  be written as follows
\be \partial_j L_v = \nabla_i S^i_j - B^i_{jk} S^k_i + \nabla^i
(H^k_{il}  H^l_{jk} )  \label{44},\ee
where
$$ S^i_j = \nabla_k F^{ki}_j  + B^i_{kl} F^{kl}_j - B^l_{kj} F^{ki}_l =
 E^i_{\mu} \frac{\delta L_v}{\delta {E^j_{\mu}}} . $$
It is necessary to illuminate the important points under the
derivation of the identity (44).  We have $$\partial_j L_v = F^{ik}_l
(\nabla_j E^l_{\mu}) \nabla_i E^{\mu}_k + F^{ik}_l E^l_{\mu} \nabla_j
\nabla_i E^{\mu}_k. $$ With Ricci's identity  $$\nabla_j \nabla_i
E^{\mu}_k  = \nabla_i \nabla_j E^{\mu}_k - R_{jik}{^l}E^{\mu}_l $$ we can
represent  the second term in the right hand side of the starting
relation in the following form  $$ \nabla_i (F^{ik}_l E^l_{\mu} \nabla_j
E^{\mu}_k) -(\nabla_i (F^{ik}_l E^l_{\mu})) \nabla_j E^{\mu}_k - F^{ik}_l
R_{jik}{^l}.  $$ For the further transformations one needs to use identity
$$ F^{ik}_l R_{jik}{^l} = \nabla_i \nabla_k F^{ik}_j. $$ 

For Dirac's Lagrangian one can derive the
identity $$ \partial_j L_D = D_j {\bar \psi} \frac{\delta
L_D}{\delta {\bar \psi}} - \frac{\delta L_D}{\delta {\psi}} D_j {\bar \psi}
+ \frac{1}{\epsilon}(\nabla_i W^i_j - B^i_{jk} W^k_i +e F_{ji} J^i) .$$
From this identity it follows in accordance with (39) and (40) that the
circulation of the energy of the spinning matter is defined by the
equation\be \nabla_i W^i_j - B^i_{jk} W^k_i + qF_{ji} J^i = 0
\label{45},\ee when the electromagnetic and vacuum fields are
present. The canonical energy-momentum tensor  $W^i_j $ of the spinning
matter is not symmetric.

\section{Conclusion}
Here we suggest an experiment to test the  formulated theory.
It is suggested to measure  the gravitational acceleration of electrons and
positrons in the Earth gravitational field. The motivation is as follows.

In 1967 Witteborn and Fairbank measured the net vertical component of
gravitational force on electrons in vacuum enclosed by a copper tube 
~\cite{5}.
This force was shown to be less than 0.09 mg, where m is the inertial mass
of the electron and g is $980 cm/sec^2.$ They concluded
that this result supports the contention that gravity induces an electric
field outside a metal surface, of such magnitude and direction  that
the gravitational force on electrons is cancelled. If this is true, then
the positrons will fall in this tube with the acceleration $a = 2g.$ The
conclusion from the theory presented here is that electrons and positrons
do not interact with the gravitational field directly but only through the
vacuum field and electromagnetic channel.  And the result presented by 
the measurements may be
considered as an estimation for the energy of vacuum field generated by
electron (and positron). Thus, the new measurements  of the net vertical
component of the force on positrons in vacuum enclosed by a copper tube
will have the fundamental significance for understanding of the conceptual
basis of contemporary theoretical physics and for the understanding 
of the nature of dark matter as well.

\end{document}